\documentclass[10pt]{article}
\usepackage{array}
\usepackage{amsmath}
\usepackage{amssymb}
\usepackage{url}

\usepackage{graphicx}

\usepackage{cite}

\usepackage{color}

\usepackage{epstopdf}

\usepackage{subfigure}
\RequirePackage{lineno}

\usepackage{graphicx}  
\usepackage{dcolumn}   
\usepackage{bm}        
\usepackage{amssymb}   

\usepackage[labelfont=bf,labelsep=period,justification=raggedright]{caption}

\makeatletter
\renewcommand{\@biblabel}[1]{\quad#1.}
\makeatother
\usepackage{tabularx}

\date{}

\begin{document}

\modulolinenumbers[1]
\begin{flushleft}
{\Large
\textbf{A hierarchical network approach for modeling Rift Valley fever epidemics with applications in North America}
}
\\
Ling Xue $^{1}$,
Lee W. Cohnstaedt $^{2,\ast}$,
H. Morgan Scott $^{3}$,
Caterina Scoglio $^{1}$
\\
\bf{1}  Kansas State Epicenter, Department of Electrical \&  Computer Engineering, Kansas State University, \ U.S. \ 66506,
\bf{2}  Center for Grain and Animal Health Research, United States Department of Agriculture, \ U.S.  \ 66502,
\bf{3} Department of Diagnostic Medicine/Pathobiology, Kansas State University, \ U.S. \ 66506
\\
$\ast$ E-mail: lee.cohnstaedt@ars.usda.gov
\end{flushleft}

\section*{Abstract}
Rift Valley fever is a vector-borne zoonotic disease which causes high morbidity and mortality in livestock. In the event Rift Valley fever virus is introduced to the United States or other non-endemic areas, understanding the potential patterns of spread and the areas at risk based on disease vectors and hosts will be vital for developing mitigation strategies. Presented here is a general network-based mathematical model of Rift Valley fever. Given a lack of empirical data on disease vector species and their vector competence, this discrete time epidemic  model uses stochastic parameters following several PERT distributions to model the dynamic interactions between hosts and likely North American mosquito vectors in dispersed geographic areas. Spatial effects and climate factors are also addressed in the model. The model is applied to a large directed asymmetric network of $3,621$ nodes based on actual farms to examine a hypothetical introduction to some counties of Texas, an important ranching area in the United States of America (U.S.A.). The nodes of the networks represent livestock farms, livestock markets, and feedlots, and the links represent cattle movements and mosquito diffusion between different nodes. Cattle and mosquito (\it Aedes \rm and \it Culex\rm) populations are treated with different contact networks to assess virus propagation. Rift Valley fever virus spread is assessed under various initial infection conditions (infected mosquito eggs, adults or cattle).  A surprising trend is fewer initial infectious organisms result in a longer delay before a larger and more prolonged outbreak. The delay is likely caused by a lack of herd immunity while the infections expands geographically before becoming an epidemic involving many dispersed farms and animals almost simultaneously. Cattle movement between farms is a large driver of virus expansion,  thus quarantines can be efficient mitigation strategy to prevent further geographic spread.

\section*{Introduction}
\label{sec:introduction}
Rift Valley fever (RVF) was first identified in Egypt in $1931$ \cite{Daubney1931} and is endemic in the eastern and southern regions of Africa \cite{favier2006rift}. Viral infection may result in abortion in adults and death in newborn livestock \cite{Martin2008impact}. Sheep, goats and cattle are the most important domestic animal hosts affected when viewed from an economic standpoint \cite{favier2006rift} although humans also can become infected \cite{Chevalier2010, Martin2008impact}.

\it Aedes \rm and \it Culex \rm genera of mosquitoes are thought to be  main RVF disease vectors with respect to vector competence \cite{Chevalier2010}. The virus is maintained between epidemics through vertical transmission within the \it Aedes \rm mosquitoes \cite{Linthicum1985}, and is thought to be propagated and amplified during epidemics by both \it Aedes \rm and \it Culex \rm species mosquitoes. High RVF transmission is typically related to persistent, above average rainfall and El Ni\~{n}o/Southern Oscillation (ENSO) events in Eastern Africa which create favorable mosquito habitats \cite{Linthicum1999}. \it Aedes \rm mosquitoes lay eggs in dry mud \cite{Zeller1997} and the eggs can survive for long periods of time \cite{favier2006rift}. After flooding, RVF virus-infected eggs can develop into infected adult mosquitoes \cite{favier2006rift}. Infected adult \it Aedes \rm mosquitoes then feed on animals which become infected, and spread the infection to other \it Aedes \rm and \it Culex \rm genera adult mosquitoes feeding on  infected animals.

 Animal movements, typically  motivated by livestock trading and marketing may accelerate the transmission of zoonotic  diseases among animal holdings which may cover a vast area \cite{Bajardi2011}. In $1977$, the trade of sheep from east Africa during Ramadan was considered to be a likely pathway for the introduction of RVF virus to Egypt \cite{Sellers1982, abdo2010risk, Davies2006}. A boy from Anjouan, an island of Comoros archipelago, was diagnosed to have been infected with RVF virus on the French island of Mayotte in  $2007$ \cite{Chevalier2010}.  The Rift Valley fever virus was likely to be  introduced by live ruminants imported from Kenya or Tanzania in the trade during the 2006-2007 Rift Valley fever outbreak \cite{Chevalier2010}.

Humans can acquire the infection from the bites of infected mosquitoes or directly from contact with the bodily fluids of infected animals \cite{EFSA2005}. Individuals working with animals, such as farmers and veterinarians, are the most vulnerable to RVF virus infection during animal outbreaks\cite{NICD2010} because of increased exposure to mosquitoes in an outdoor environment and direct contact with animals. Rift Valley fever virus infection causes severe influenza-like disease in humans with serious consequences such as blindness, or even death \cite{Martin2008impact}. It has been reported that more than $200$ persons died of RVF in Mauritania in $1987$ \cite{Jouan1988}. There were $738$ reported human cases in Sudan, including $230$ deaths, in $2007-2008$ \cite{WHO2007}. It is likely that the number of human cases has been underreported in the past, especially in rural areas \cite{Chevalier2010}. Rift Valley fever virus has spread outside of Africa to Yemen and Saudi Arabia in $2000$ \cite{Chevalier2010} and the French island of Mayotte with multiple human cases reported \cite{sissoko2009rift}. Rift Valley fever virus could possibly be introduced to the United States, similar to the experience with West Nile virus which was introduced into the North America in $1999$ \cite{Kilpatrick2011}. A mathematical epidemiological model can be applied to non-traditional locations in order to study the potential for spatial spread of RVF virus.

Epidemiological modeling plays an important role in planning, implementing, and evaluating detection, control, and prevention programs \cite{Ma2009}. Mathematical modeling takes the advantage of economic, clear and precise mathematical formulation, e.g.,  applications of differential, integral, or functional differential equations \cite{Ma2009}. Mathematical models of infection transmission include interpretation of transmission processes and are often useful in answering questions that cannot be answered only with empirical data analysis \cite{metras2011rift}, as well as to explore biological and critical ecological characteristics of disease transmission \cite{Ross1916, Luz2010}. Current RVF virus transmission models are useful in representing infection transmission process \cite{metras2011rift} but are limited in determining and testing relevant risk factors. For the Ferlo area of Senegal, a pond-level meta-population model which considered only vectors was developed assuming that \it Aedes \rm mosquitoes were the only vector and rainfall was the only driving force \cite{favier2006rift}. It has been shown that within Ferlo, the virus would persist only if the livestock moved between ponds and the rainfall did not occur in all ponds simultaneously \cite{favier2006rift}. Very few mathematical dynamic transmission models have explored mechanisms of RVF virus circulation \cite{metras2011rift} on a larger geographical scale. A theoretical model in a closed system including \it Aedes \rm and \it Culex \rm mosquitoes and livestock population was earlier proposed \cite{gaff2007epidemiological}. The key result was that RVF virus can persist in a closed system for $10$ years if the contact rate between hosts and vectors is high \cite{metras2011rift, gaff2007epidemiological}. Another theoretical RVF virus transmission mathematical model \cite{ Mpeshe2011}  modified the model in \cite{gaff2007epidemiological} by adding human hosts, merging all mosquitoes into one class, removing mosquito egg compartment, as well as vertical transmission of mosquitoes.   Sensitivity indices of the reproduction number are used to determine the most sensitive parameters to the basic reproduction number of RVF virus transmission \cite{ Mpeshe2011}. It has been found that both the reproduction number and disease prevalence in mosquitoes are sensitive to mosquito death rate and the disease prevalence in livestock and humans are more sensitive to livestock and human recruitment rates \cite{Mpeshe2011}.  A theoretical ordinary differential equation  meta-population involving livestock and human mobility was presented \cite{Niu2012}. They analyzed the likelihood of pathogen establishment and provided hypothesized examples to illustrate the methodology \cite{Niu2012}. A three-patch model for the process by which animals enter Egypt from Sudan, are moved up the Nile, and then consumed at population centers is proposed \cite{Gao2013}.    Using \cite{gaff2007epidemiological} and \cite{ Mpeshe2011} as a foundation, the homogeneous models have been extended to a meta-population  differential equation model   including \it Aedes\rm, \it Culex\rm, livestock, and humans and a case study was carried out for South Africa during a country-wide outbreak in 2010 \cite{LingXue2010}.  The model  was based on RVF virus spatial transmission during an outbreak, where a network with three nodes corresponding to three affected provinces in South Africa was established. To make the output of the model \cite{LingXue2010}  easily compared with incidence data if available and the simulation for thousands of nodes easily  implemented, a discrete time epidemic  model is developed and a much larger network on which to study the dynamics of the larger system is established.

Proposed here is a deterministic network-based RVF virus transmission model with stochastic parameters. Two competent vector populations: \it Aedes \rm mosquitoes, \it Culex \rm mosquitoes, and two host populations: cattle and humans are considered. The dynamical behavior of mosquito and livestock    populations are  modeled  using  a meta-population approach  based on weighted contact networks. The nodes of the networks represent geographical locations, and the weights represent the level of contact between regional pairings. In particular, nodes represent different farm sizes or operator businesses of the cattle industry, nominally markets and feedlots. Heterogeneous aspects of the spreading are considered in the model through realistic modeling of the cattle movement among different nodes of the network. Additionally, the mosquito population and development is modeled as a function of climatic factors, such as humidity and temperature. It is easy to implement simulations of  the model even for networks with thousands of nodes, and it is easy to compare the output of the model with incidence data if available. The role of starting location has been shown to be important in the final size of  rinderpest epidemic \cite{Manore2011}. To investigate the role of starting location, and the size of initial infection in RVF virus spread,  the proposed model has been applied to a case study to some counties  in Texas, U.S. and the model outcomes (the human and cattle cases, and the timing of the epidemic's characteristics) indicate which biotic factors will play an important role if RVF virus is introduced to the United States.

\allowdisplaybreaks
\section*{Methods and Materials}
\label{sec:MaterialsandMethods}

\subsection*{Network-based Meta-population Models}
\begin{figure}[!htbp]
\centering
\subfigure[An example of mosquito diffusion network]{
\label{fig:mosquitonetwork} 
\includegraphics[angle=0,width=10cm,height=8cm]{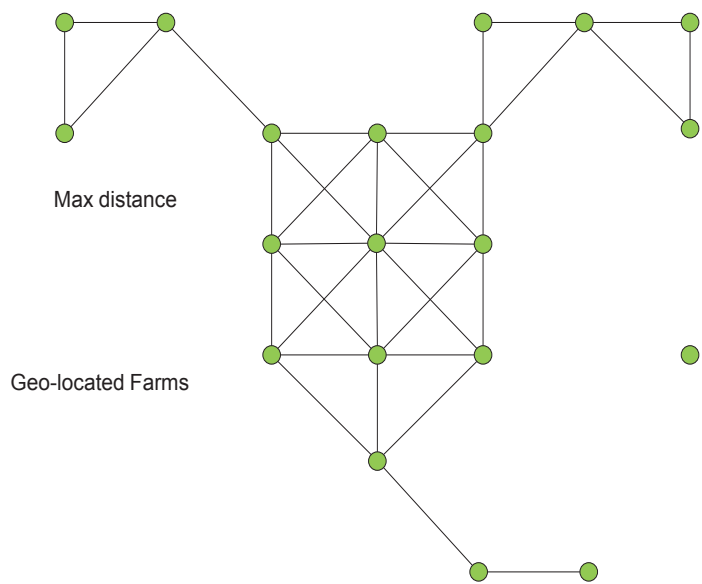}}
\subfigure[An example of livestock movement network]{
\label{fig:cattlenetwork} 
\includegraphics[angle=0,width=10cm,height=8cm]{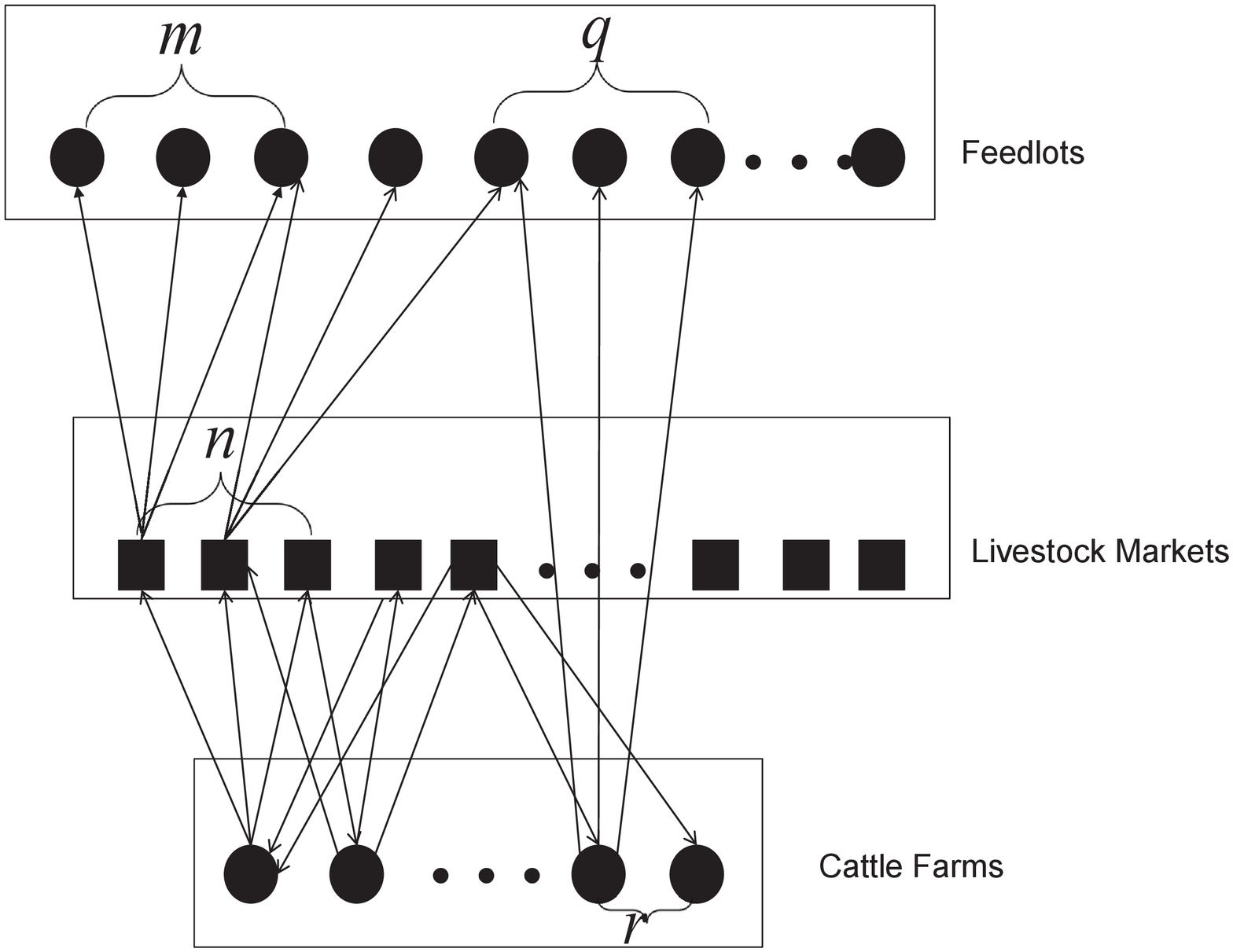}}
\hspace{0.1in}
\caption{Mosquito diffusion and livestock movement illustration}
\label{fig:diagramandmosnetwork} 
\end{figure}

\it Aedes \rm mosquitoes, \it Culex \rm mosquitoes, livestock, and human populations each are considered in the network-based meta-population models. The movement of each population is represented by networks, where nodes denote locations, and links denote movement flow between locations. In the mosquito diffusion network, the nodes represent farms and the links represent mosquito diffusion from one farm to the neighboring farms. The weights are diffusion rates $\omega_{1ij}$ for \it Aedes \rm population, and $\omega_{3ij}$ for \it Culex \rm population from location $i$ to location $j$. In the livestock movement network, the nodes represent farms, livestock markets, and feedlots. The links represent livestock movements due to livestock trade between the nodes and the weight is the movement rate $\omega_{2ij}$ from node $i$ to node $j$. The mosquito and livestock networks are shown in Fig. \ref {fig:mosquitonetwork} and Fig. \ref{fig:cattlenetwork}, respectively.

The compartmental models are adapted to represent the status of each population during a simulated  RVF virus transmission. The models are built  based on the principle of the RVF virus transmission flow diagram illustrated in \cite{LingXue2010}. Adult \it Aedes \rm and \it Culex \rm populations are distributed among susceptible $S_{ai}$, exposed $E_{ai}$, and infected $I_{ai}$ compartments. Only those mosquito species that are known to be competent vectors of RVF virus transmission are considered  and they are broadly grouped by \it Aedes \rm and \it Culex \rm genera mosquitoes. The subscript $a=1$ denotes \it Aedes \rm in node $i$, and $a=3$  denotes \it Culex \rm mosquitoes in node $i$. Uninfected and infected mosquitoes eggs are represented by $P_{ai}$ and $Q_{ai}$, respectively.  \it Culex \rm mosquitoes do not display vertical transmission. Therefore, only uninfected \it Culex \rm eggs are incorporated in the model. The livestock and human hosts are likewise considered $S_{bi}$, $E_{bi}$, $I_{bi}$, and $R_{bi}$. The subscript representing livestock in node $i$ is $b=2$, and humans in node $i$ are represented with $b=4$. The descriptions of the parameters in the models are found in Table \ref{table:parameters}. All the transitions to be discussed below are for location $i$ at day $t$.
\subsubsection*{\it Aedes \rm  \bf{Population Model}}
\begin{align}
  P_{1i}(t+1)- P_{1i}(t) &= b_1(N_{1i}(t)-q_1I_{1i}(t)) -\theta_1P_{1i}(t)) \label{equation:Aedeseggs}\\
  Q_{1i}(t+1)- Q_{1i}(t)&=b_1 q_1I_{1i}(t)-\theta_1Q_{1i}(t)\\
 S_{1i}(t+1)- S_{1i}(t) &=\theta_1P_{1i}(t)+\sum^n_{j=1, j \neq i}\omega_{1ji}S_{1j}(t)-\sum^n_{j=1, j \neq i}\omega_{1ij}S_{1i}(t)-d_1S_{1i}(t)N_{1i}(t)/K_1\nonumber\\&-\beta_{21}S_{1i}(t)I_{2i}(t)/N_{2i}(t)\\
   E_{1i}(t+1)- E_{1i}(t) &=\sum^n_{j=1, j \neq i}\omega_{1ji}E_{1j}(t)-\sum^n_{j=1, j \neq i}\omega_{1ij}E_{1i}(t)-d_1E_{1i}(t)N_{1i}(t)/K_1+\beta_{21}S_{1i}(t)I_{2i}(t)/N_{2i}(t)\nonumber\\&-\varepsilon_1E_{1i}(t)\\
I_{1i}(t+1)- I_{1i}(t)&=\sum^n_{j=1, j \neq i}\omega_{1ji}I_{1j}(t)-\sum^n_{j=1, j \neq i}\omega_{1ij}I_{1i}(t)+\theta_1Q_{1i}(t)-d_1I_{1i}(t)N_{1i}(t)/K_1\nonumber\\&+\varepsilon_1E_{1i}(t)\\
N_{1i}(t+1)&=S_{1i}(t+1)+E_{1i}(t+1)+I_{1i}(t+1)
\end{align}

There are $b_1N_{1i}(t)$ eggs laid, including $ b_1 q_1I_{1i}(t)$ infected eggs, and $b_1N_{1i}(t)-b_1 q_1I_{1i}(t)$ uninfected eggs each day. After the development period, $\theta_1P_{1i}(t)$ uninfected eggs develop into susceptible adult \it Aedes \rm mosquitoes and $\theta_1Q_{1i}(t)$ infected eggs develop into infected adult \it Aedes \rm mosquitoes. The number of \it Aedes \rm mosquitoes infected  by  livestock is $\beta_{21}S_{1i}(t)I_{2i}(t)/N_{2i}(t)$. Following the incubation period, $\varepsilon_1E_{1i}(t)$ \it Aedes \rm mosquitoes transfer from exposed compartment to infected compartment. The number of \it Aedes \rm mosquitoes dying naturally in compartment $X$ is given as $d_1X_{1i}(t)$. The percentage of \it Aedes \rm mosquitoes moving from location $i$ to location $j$ is $\omega_{1ij}$. The change in the number of \it Aedes \rm mosquitoes due to mobility in compartment $X$ is given as $\sum^n_{j=1, j \neq i}\omega_{1ji}X_{1j}(t)-\sum^n_{j=1, j \neq i}\omega_{1ij}X_{1i}(t)$ \cite{Keeling2008}.

\subsubsection*{\it Culex \rm \textbf{Population Model}}
\allowdisplaybreaks
\begin{align}
  P_{3i}(t+1)- P_{3i}(t) &=b_3(t)N_{3i}(t)-\theta_3(t)P_{3i}(t)\\
  S_{3i}(t+1)- S_{3i}(t)  &=\theta_3(t)P_{3i}(t)+\sum^n_{j=1, j \neq i}\omega_{3ji}S_{3j}(t)-\sum^n_{j=1, j \neq i}\omega_{3ij}S_{3i}(t)-d_3S_{3i}(t)N_{3i}(t)/K_3\nonumber\\&-\beta_{23}S_{3i}(t)I_{2i}(t)/N_{2i}(t)\\
   E_{3i}(t+1)- E_{3i}(t) &=\sum^n_{j=1, j \neq i}\omega_{3ji}E_{3j}(t)-\sum^n_{j=1, j \neq i}\omega_{3ij}E_{3i}(t)-\varepsilon_3E_{3i}(t)-d_3E_{3i}(t)N_{3i}(t)/K_3\nonumber\\&+\beta_{23}S_{3i}(t)I_{2i}(t)/N_{2i}(t)\\
   I_{3i}(t+1)- I_{3i}(t)  &=\sum^n_{j=1, j \neq i}\omega_{3ji}I_{3j}(t)-\sum^n_{j=1, j \neq i}\omega_{3ij}I_{3i}(t)+\varepsilon_3E_{3i}(t)-d_3I_{3i}(t)N_{3i}(t)/K_3\\
N_{3i}(t+1)&=S_{3i}(t+1)+E_{3i}(t+1)+I_{3i}(t+1)
\end{align}
There are $b_3N_{3i}(t)$ eggs laid each day. After the development period, $\theta_3P_{3i}(t)$ eggs develop into susceptible adult \it Culex \rm mosquitoes. After the incubation period, $\varepsilon_3E_{3i}(t)$ \it Culex \rm mosquitoes transfer to infected compartment $I$. The number of \it Culex \rm mosquitoes acquiring infection from livestock is denoted by $\beta_{23}S_{3i}(t)I_{2i}(t)/N_{2i}(t)$. The \it Culex \rm mosquitoes removed from compartment $X$ due to natural death is $d_3X_{3i}(t)$. The percentage of \it Culex \rm mosquitoes moving from location $i$ to location $j$ is $\omega_{3ij}$. The change in the number of \it Culex \rm mosquitoes due to movement in compartment $X$ is given as $\sum^n_{j=1, j \neq i}\omega_{3ji}X_{3j}(t)-\sum^n_{j=1, j \neq i}\omega_{3ij}X_{3i}(t)$ \cite{Keeling2008}.

 \subsubsection*{Livestock Population Model}

\allowdisplaybreaks
\begin{align}
 S_{2i}(t+1)- S_{2i}(t)  &=b_2(t)\delta_b (i)N_{2i}(t)+\sum^n_{j=1, j \neq
i}\omega_{2ji}S_{2j}(t)-\sum^n_{j=1, j \neq
i}\omega_{2ij}S_{2i}(t)-d_2\delta_d(i)S_{2i}(t)N_{2i}(t)/K_{2}\nonumber\\&-\beta_{12}S_{2i}(t)I_{1i}(t)/N_{1i}(t)
-\beta_{32}S_{2i}(t)I_{3i}(t)/N_{3i}(t) \\
E_{2i}(t+1)- E_{2i}(t)  &=\sum^n_{j=1, j \neq i}\omega_{2ji}E_{2j}(t)-\sum^n_{j=1, j \neq i}\omega_{2ij}E_{2i}(t)-d_2\delta_d (i)E_{2i}(t)N_{2i}(t)/K_2-\varepsilon_2E_{2i}(t)\nonumber\\&+\beta_{12}S_{2i}(t)I_{1i}(t)/N_{1i}(t)
+\beta_{32}S_{2i}(t)I_{3i}(t)/N_{3i}(t)\\
I_{2i}(t+1)- I_{2i}(t) &=p\sum^n_{j=1, j \neq i}\omega_{2ji}I_{2j}(t)-p\sum^n_{j=1, j \neq i}\omega_{2ij}I_{2i}(t)-d_2\delta _d(i)I_{2i}(t)N_{2i}(t)/K_2+\varepsilon_2E_{2i}(t)\nonumber\nonumber\\&-\gamma_2I_{2i}(t)-\mu_2I_{2i}(t)\\
R_{2i}(t+1)- R_{2i}(t) &=\sum^n_{j=1, j \neq i}\omega_{2ji}R_{2j}(t)-\sum^n_{j=1, j \neq i}\omega_{2ij}R_{2i}(t)+\gamma_2I_{2i}(t)-d_2\delta_d(i)R_{2i}(t)N_{2i}(t)/K_2\\
N_{2i}(t+1)&=S_{2i}(t+1)+E_{2i}(t+1)+I_{2i}(t+1)+R_{2i}(t+1)
\end{align}

The daily number of newborn livestock in location $i$ is $ b_2(t)N_{2i}(t)$. The variables  $\delta_b (i)$ and  $\delta_d (i)$ are used to differentiate different types of nodes. If location $i$ is a farm, then $\delta_b (i)=1$, $\delta_d (i)=1$. If location $i$ is a market, then $\delta_b (i)=0$, $\delta_d (i)=0$. If location $i$ is a feedlot, then $\delta_b (i)=0$, $\delta_d (i)=1$. The numbers of livestock infected by \it Aedes \rm mosquitoes and \it Culex \rm mosquitoes are denoted by $\beta_{12}S_{2i}(t)I_{1i}(t)/N_{1i}(t)$ and $\beta_{32}S_{2i}(t)I_{3i}(t)/N_{3i}(t)$, respectively. After the incubation period, $\varepsilon_2E_{2i}(t)$ livestock transfer from exposed state to infected state. After the infection period, $\gamma_2I_{2i}(t)$ livestock recover from RVF virus infection. The number of dead livestock in  compartment $X$ is given as $d_2X_{2i}N_{2i}(t)/K_2$ in which $K_2$ is the carrying capacity of livestock in each node. The change in the number of livestock in compartment $X$ due to mobility is given as $\sum^n_{j=1, j \neq i}\omega_{2ji}X_{2j}(t)-\sum^n_{j=1, j \neq i}\omega_{2ij}X_{2i}(t)$ for livestock in compartments $S$, $E$, and $R$,  and $p\sum^n_{j=1, j \neq i}\omega_{2ji}X^{[m]}_{2j}(t)$-$p\sum^n_{j=1, j \neq i}\omega_{2ij}X^{[m]}_{2i}(t)$ \cite{Keeling2008}, $(0<p<1)$ for livestock in compartment $I$.
\subsubsection*{Human Population Model}

\allowdisplaybreaks
\begin{align}
S_{4i}(t+1)- S_{4i}(t)&=-\beta_{14}S_{4i}(t)I_{1i}(t)/N_{1i}(t)-\beta_{24}S_{4i}(t)I_{2i}(t)/N_{2i}(t)-
\beta_{34}S_{4i}(t)I_{3i}(t)/N_{3i}(t)\\
E_{4i}(t+1)- E_{4i}(t)&=\beta_{14}S_{4i}(t)I_{1i}(t)/N_{1i}(t)+\beta_{24}S_{4i}(t)I_{2i}(t)/N_{2i}(t)+
\beta_{34}S_{4i}(t)I_{3i}(t)/N_{3i}(t)\nonumber\\&-\varepsilon_4E_{4i}(t)\\
 I_{4i}(t+1)- I_{4i}(t)  &=\varepsilon_4E_{4i}(t)-\gamma_4I_{4i}(t)\\
R_{4i}(t+1)- R_{4i}(t) &=\gamma_4I_{4i}(t) \label{equation:recoveredhuman}
\end{align}

The number of humans in each node is constant because  birth, death, mortality, and mobility of humans are not considered. The number of humans infected by \it {Aedes} \rm mosquitoes, \it Culex \rm mosquitoes, and livestock is $\beta_{14}S_{4i}(t)I_{1i}(t)/N_{1i}(t)$, $\beta_{24}S_{4i}(t)I_{2i}(t)/N_{2i}(t)$, and $\beta_{34}S_{4i}(t)I_{3i}(t)/N_{3i}(t)$, respectively. There are $\varepsilon_4E_{4i}(t)$ humans transferring to  infected compartment after incubation period, and $\gamma_4I_{4i}(t)$ humans recovering from RVF virus infection after infection period.
\begin{table}[!ht]
\caption{Parameter ranges for numerical simulations.}
\centering
\begin{tabular}{|p{45pt}p{170pt} p{80pt} p{40pt} p{25pt} p{60pt}|}
\hline
\textbf{Parameter} & \textbf{Description} &\textbf{Range}& \textbf{Assumed most possible value}& \textbf{Units} & \textbf{Source}\\
\hline
$\beta_{12}$ & contact rate: \it Aedes \rm to livestock & $(0.0021, 0.2762)$& 0.1392&$1/$day &\cite{Canyon1999, Hayes1973, Jones1985, Magnarelli1977, PrattMoore1993, Turell1988, Turell1988b}\\
$\beta_{21}$ & contact rate: livestock to \it Aedes \rm & $(0.0021, 0.2429)$ & 0.1225&$1/$day &\cite{Canyon1999, Hayes1973, Jones1985, Magnarelli1977, PrattMoore1993, Turell1987}\\
$\beta_{23}$ & contact rate: livestock to \it Culex \rm &$(0.0000, 0.3200)$ & 0.16 &$1/$day &\cite{Hayes1973, Jones1985, Magnarelli1977, PrattMoore1993, Turell1987, Wekes1997}\\
$\beta_{32}$ &contact rate: \it Culex \rm to livestock &$ (0.0000, 0.096 )$ & 0.04&$1/$day & \cite{Hayes1973, Jones1985, Magnarelli1977, PrattMoore1993, Wekes1997}\\
$\beta_{14}$ & contact rate: \it Aedes \rm to humans & $(0.001, 0.002)$& 0.0015 &$1/$day & Assume\\
$\beta_{24}$& contact rate: livestock to humans &$(0.00004, 0.00008)$&0.00006 &$1/$day &Assume\\
$\beta_{34}$& contact rate: \it Culex \rm to humans & $(0.0005,0.001)$ & 0.000525 &$1/$day &Assume\\
$1/\gamma_2$& recovery period in livestock &$(2, 5) $&3.5 &$1/$day &\cite{Erasmus1981}\\
$1/\gamma_4$ &recovery period in humans &$(4, 7)$ & 5.5&$1/$day & \cite{Mpeshe2011}\\
$1/d_1$ & longevity of \it Aedes \rm mosquitoes &$ (3, 60)$ & 31.5 &days &\cite{Bates1970, Moore1993, PrattMoore1993}\\
$1/d_2$ & longevity of livestock &$(360, 3600)$& 1980 &days & \cite{Radostits2001}\\
$1/d_3$ & longevity of \it Culex \rm mosquitoes &$(3, 60)$& 31.5 &days & \cite{Bates1970, Moore1993, PrattMoore1993}\\
$b_1$ & birth rate of \it Aedes \rm mosquitoes &weather dependent &  &$1/$day &\cite{Bates1970, Moore1993, PrattMoore1993}\\
$b_2$ & birth rate of livestock &$d_2$&  &$1/$day & \cite{Radostits2001}\\
$b_3$ & birth rate of \it Culex \rm mosquitoes &weather dependent &  &$1/$day & \cite{Bates1970, Moore1993, PrattMoore1993}\\
$1/\epsilon_1$ &incubation period in \it Aedes \rm mosquitoes &$ (4, 8)$&6 &days &\cite{Turell1988}\\
$1/\epsilon_2$ &incubation period in livestock &$(2, 6) $& 4 &days &\cite{Peters1994}\\
$1/\epsilon_3$ &incubation period in \it Culex \rm mosquitoes &$(4, 8) $& 6&days &\cite{Turell1988}\\
$1/\epsilon_4$ &incubation period in humans & $(2, 6) $& 4&days & \cite{ Mpeshe2011}\\
$\mu_2$ &mortality rate in livestock & $(0.025, 0.1)$& 0.0375 & $1/$day &\cite{Erasmus1981, Peters1994}\\
$q_1$ & transovarial transmission rate in \it Aedes \rm mosquitoes & $(0, 0.1) $& 0.05&$1/$day &\cite{Freier1987}\\
$1/\theta_1$&development period of \it Aedes \rm mosquitoes &weather dependent &  &days &\cite{PrattMoore1993 }\\
$1/\theta_3$&development period of \it Culex \rm mosquitoes& weather dependent & &days &\cite{gong2010climate}\\
$K_1$ &carrying capacity of \it Aedes \rm mosquitoes & &$100000000$ & &Assume\\
$K_2$&carrying capacity of livestock & & $1000000$ && Assume\\
$K_3$& carrying capacity of \it Culex \rm mosquitoes&& $100000000$&&Assume\\
$p$& reduction in $\omega_{2ij}$ due to infection &&$\frac{1}{2}$ &&Assume\\
\hline
\end{tabular}
\label{table:parameters}
\end{table}
\subsection*{Case Study: Texas, U.S.A from $2005$ to $2010$}
\label{sec:simulations}
\subsubsection*{Networks in the Study Area}
As a case study, various RVF virus introduction scenarios were tested using the model to determine the hypothetical model outcomes (number of livestock cases and timing of the epidemic).   Although the model accounts for their exact locations when simulating RVF virus spread, we do not report any of this information or even discuss ranches in areas smaller than county level.  The exact farms and counties are very well masked from the results.  Texas cattle ranches were selected because they have large cattle concentrations and we have aggregate survey data on cattle movements in these areas \cite{BRANDON2007}. A network with $3,526$ cattle farms \cite{2007Texascountycensus}, $3$ livestock markets \cite{2007Texascountycensus}, and $92$ cattle feedlots \cite{2007Texascountycensus}  is constructed. The cattle farms, and  livestock markets  are located in one region, and the feedlots are in another region. The location of each node is uniformly distributed in each county according to the total number of farms within each county \cite{2007Texascountycensus}. The exact location of each farm is obscured because those data are not publicly available \cite{Riley2010} due to confidentiality.  The initial number of cattle in each farm is categorized as $0-9$, $10-19$, $20-49$, $50-99$, $100-199$, $200-499$ and more than $500$ \cite{2007Texascountycensus}. The initial number of susceptible cattle in each farm or feedlot for numerical simulation is assumed according to the number of cattle in each county  in $2007$ \cite{2007Texascountycensus} and the histogram of the number of cattle \cite{2007Texascountycensus}. For cattle movement, if cattle are sold from one node to another, then there is a link between the nodes. The movement rate of cattle denoted by $\omega_{2ij}$ shown in Table \ref{table:movementrate} is estimated based on the aggregate movement rates from survey \cite{BRANDON2007} and inversely proportional to the distance between source-destination pairs.  Movement rate is the  average movement rate for all cattle at different ages,  and the movement rate of  cattle in compartment $I$ is assumed to be half the movement rate for cattle in compartments S, E, and R, namely, $p=\frac{1}{2}$

\begin{table}[!ht]
\caption{Cattle  movement rate $\omega_{2ij}$,  where
$n_m(i)$=the number of markets connected to farm $i$,
$n_f(i)$=the number of farms connected to market $i$,
$n_{ffe}(i)$=the number of feedlots connected to farm $i$,
$n_{mfe}(i)$=the number of feedlots connected to market $i$.}
\centering
\begin{tabular}{|l  l l l|} \hline
 $ \mathbf{i}  $&$  \mathbf{j}$&\textbf{Range}&\textbf{Source}\\
\hline
farm & market & $60.7\%$ /$(n_m(i)\times d_{ij})$&\cite{BRANDON2007}\\
market  & farm   & $60.7\%$ /$(n_f(i)\times d_{ij})$&\cite{BRANDON2007}\\
farm  & feedlot& $10.9\%$ /$(n_{ffe}(i)\times d_{ij})$&\cite{BRANDON2007}\\
market &feedlot & $10.9\%$ /$(n_{mfe}(i)\times d_{ij})$&\cite{BRANDON2007}\\
feedlot &farm &$0 $&\cite{BRANDON2007}\\
feedlot &market&$0$&\cite{BRANDON2007}\\
\hline
\end{tabular}
\label{table:movementrate}
\end{table}

For mosquito diffusion, if the distance between two farms is smaller than an assumed  radius, two  kilometers, then there is a link between the nodes in the network. The diffusion rates of \it Aedes \rm and \it Culex \rm mosquitoes are shown below \cite{otero2010stochastic}.
\[
\omega_{1ij} = \omega_{3ij} =\left\{
\begin{array}{l l}
0, & \quad \text{ if the nodes are disjoint}\\
diff/d_{ij}^2, & \quad \text{if two nodes share a border}\\
\end{array} \right.
\]
where $d_{ij} $is the distance between the centers of node $i$ and node $j$ \cite{otero2010stochastic} and $diff$ is a diffusion like parameter within the range $(830, 8300)m^2/$day \cite{otero2010stochastic}.

\subsubsection*{Parameters for Numerical Simulations }
Vector competence varies within and between mosquito species \cite{Turell2010}. Stochastic parameters were used to account for broad range of vector competence between \it Aedes \rm and \it Culex \rm species and individual variation within each species.  The PERT distribution has few constraints (minimum, maximum, and most likely value),  similarly to the triangular distribution applied in \cite{Wonham2006}  to  simulate West Nile virus epidemic.   In the following simulations, PERT distributions are  selected to generate stochastic parameters  with   ranges and the most likely values  listed in Table \ref{table:parameters}.  Any appropriate parameter distribution can be  adapted into the model.

\begin{figure}[!htbp]
\centering
\subfigure[The egg development rate of \it Culex \rm mosquitoes with temperature \cite{gong2010climate}.]{\label{fig:b1andb3}
\includegraphics[angle=0,width=6.1cm,height=5.9cm]{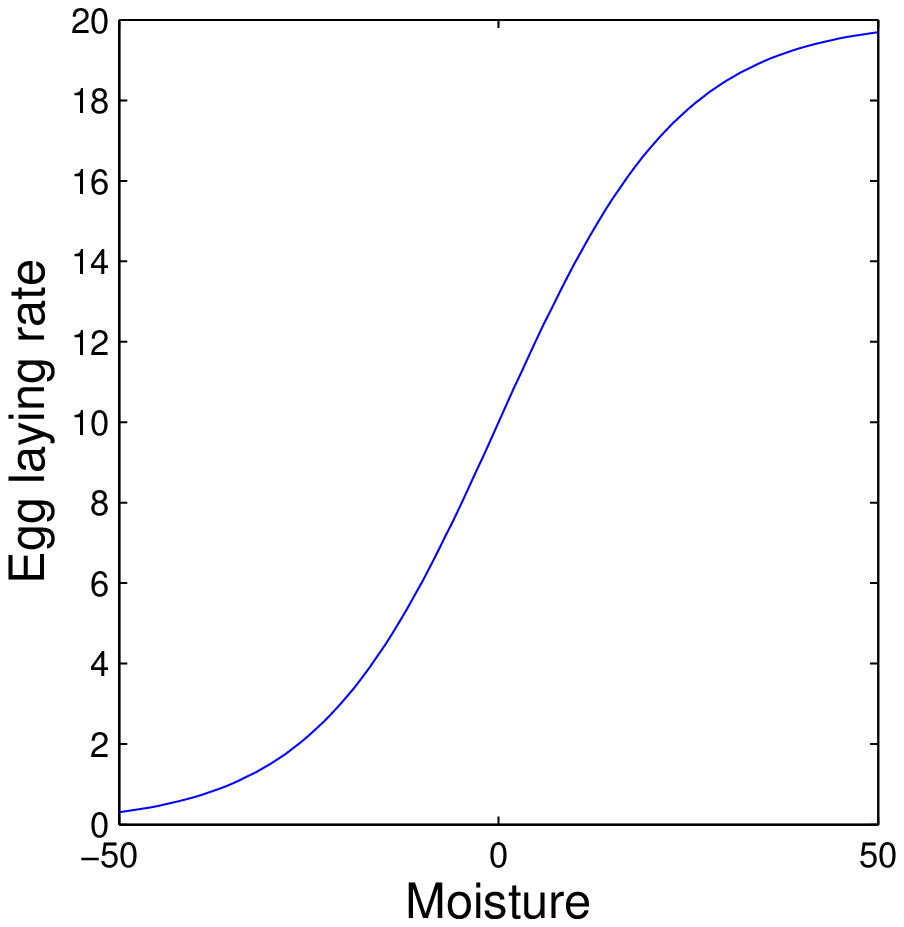}}
\hspace{0.1in}
\subfigure[ The egg development rate of \it Culex \rm mosquitoes with temperature \cite{gong2010climate}.]{\label{fig:theta3}
\includegraphics[angle=0,width=6.1cm,height=5.9cm]{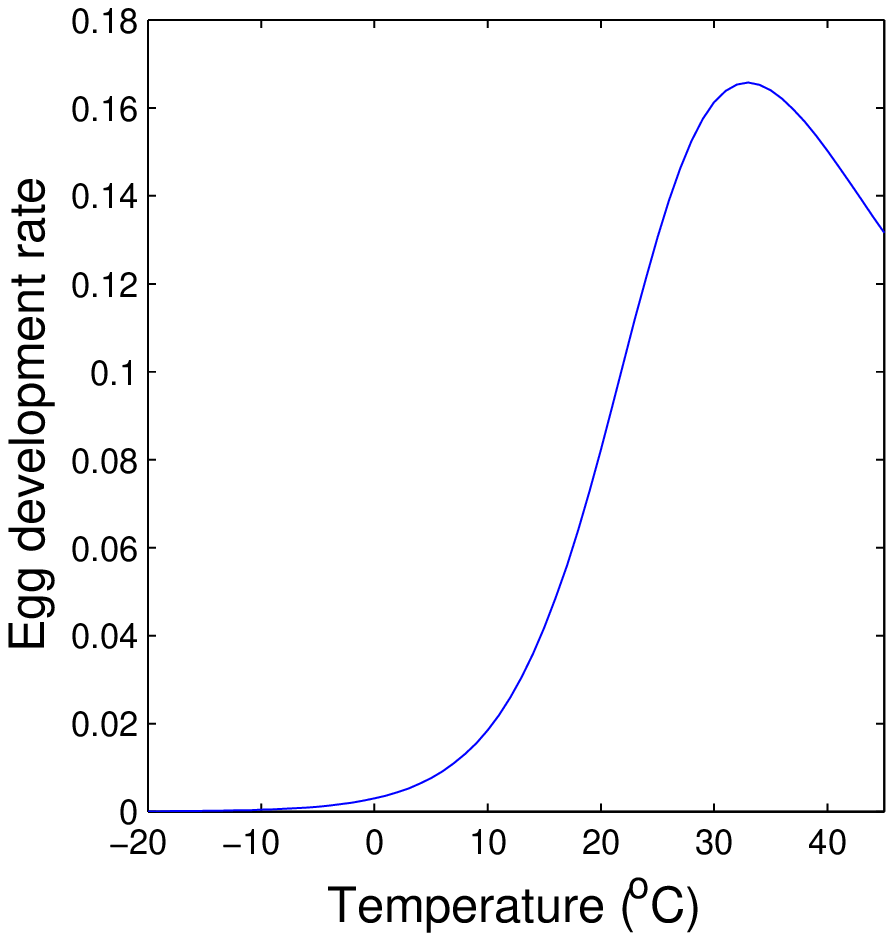}}
\hspace{0.1in}
\subfigure[The egg laying rates  of  \it Aedes \rm and \it Culex \rm  mosquitoes  in the nine counties in the south of Texas from January, $2005$ to October, $2010$.]{\label{fig:b1andb3from05to10}
\includegraphics[angle=0,width=6.1cm,height=5.9cm]{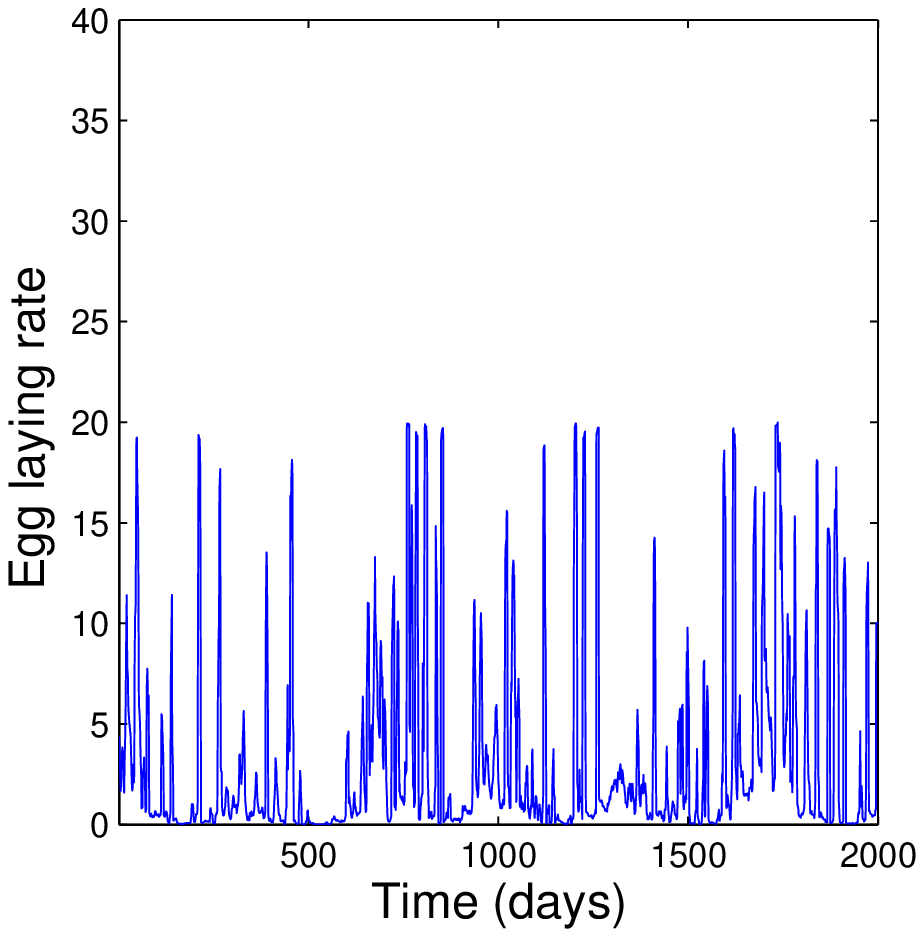}}
\hspace{0.1in}
\subfigure[The egg development rate of \it Culex \rm   mosquitoes in the nine counties in the south of Texas from January, $2005$ to October, $2010$.]{\label{fig:theta3from05to10}
\includegraphics[angle=0,width=6.1cm,height=5.9cm]{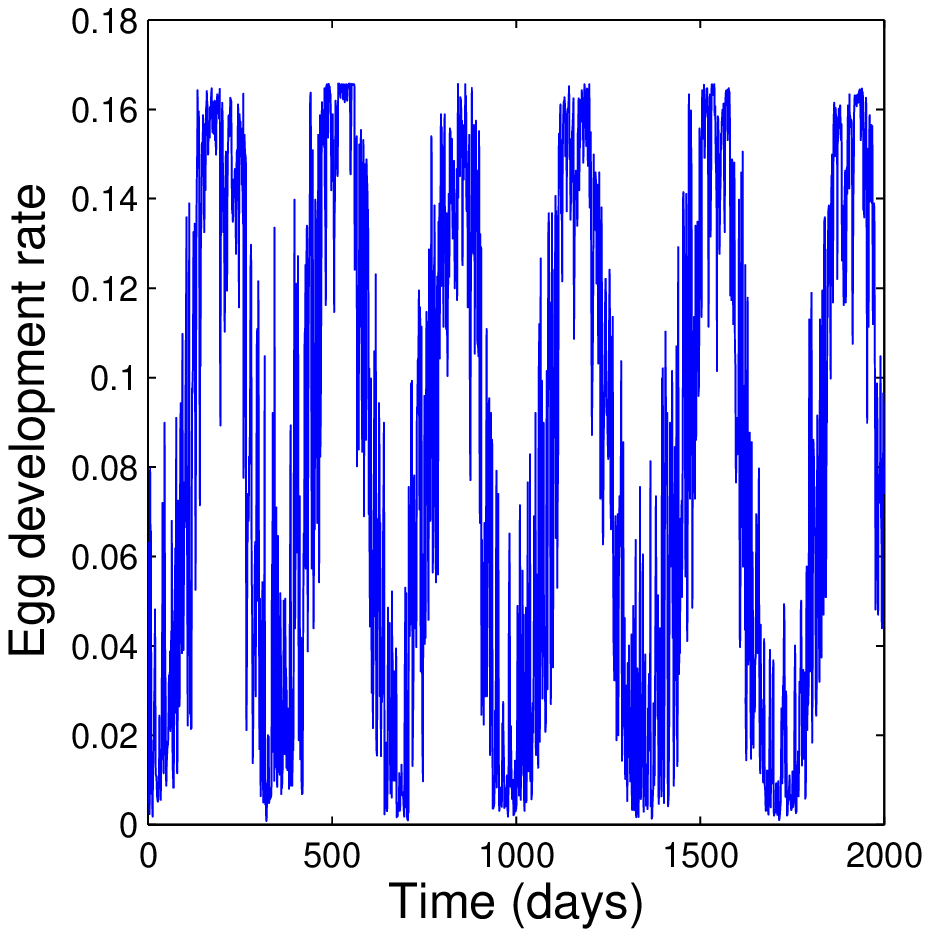}}
\hspace{0.1in}
\subfigure[ The egg development rate of \it Aedes \rm mosquitoes with temperature.]{\label{fig:theta1}
\includegraphics[angle=0,width=6.1cm,height=5.9cm]{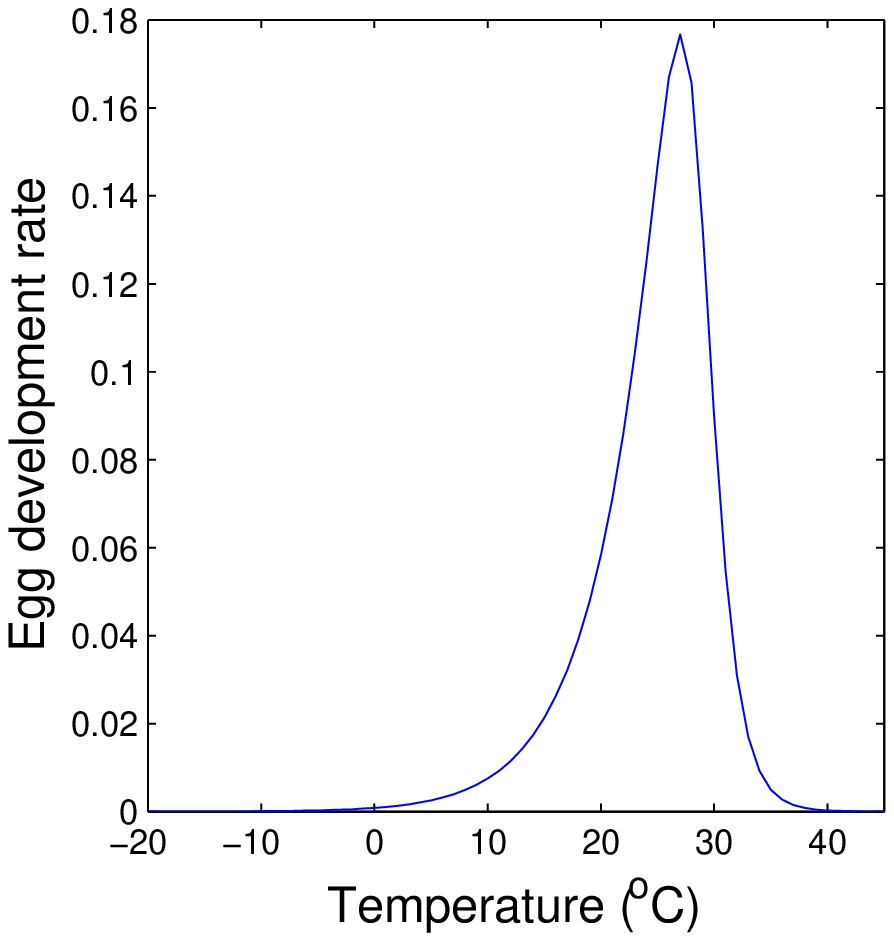}}
\hspace{0.1in}
\subfigure[The egg development rate of \it Aedes \rm   mosquitoes in the nine counties in the south of Texas from January, $2005$ to October, $2010$.]{\label{fig:theta1from05to10}
\includegraphics[angle=0,width=6.1cm,height=5.9cm]{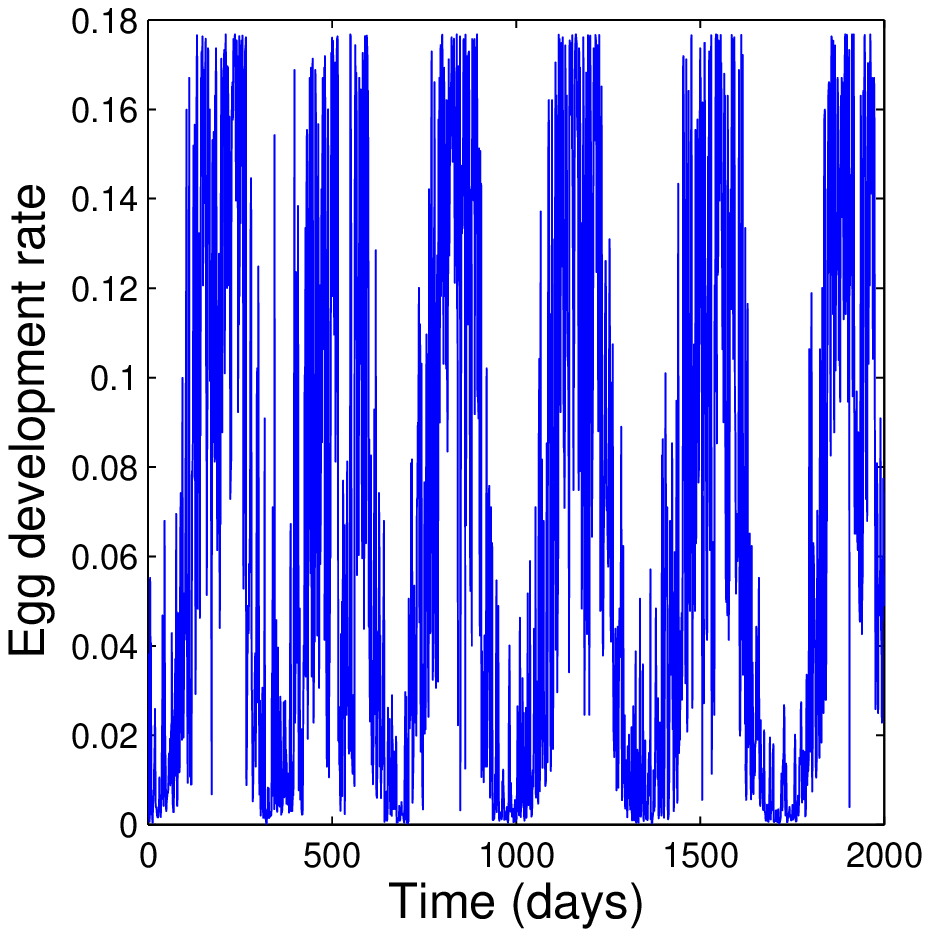}}
\hspace{0.1in}
\caption{The relationship between egg laying rates, egg development rates of mosquitoes and climate factors.}
\label{fig:mosquitoparameters} 
\end{figure}

The egg laying rates of  \it Aedes \rm  and  \it Culex \rm mosquitoes changing with moisture conditions as indicated in Equation (\ref{equation:birthrate}) \cite{gong2010climate} are shown in  Fig. \ref{fig:b1andb3}.  The egg development rate of \it Aedes \rm mosquitoes varying with temperature  in   Equation  $(\ref{equation:theta1})$ \cite{Rueda1990} and that of \it Culex \rm mosquitoes in  Equation  $(\ref{equation:developmentrate})$ are in Fig. \ref{fig:theta1} and Fig. \ref{fig:theta3}, respectively.   
 The parameters for egg laying rates of \it Aedes \rm mosquitoes  and \it Culex \rm  mosquitoes, and parameters for egg development rate of  \it Culex \rm  mosquitoes are derived from data concerning West Nile virus in $2010$ in the Northern U.S. \cite{gong2010climate}, and the parameters for the egg development rate of   \it Aedes \rm mosquitoes is derived using  the model for  \it Aedes \rm  aegypti  \cite{Rueda1990}, which are the best models currently available. More precise parameters can be adopted, as they become available. The egg laying rate of \it Aedes \rm and \it Culex \rm mosquitoes, egg development rate of \it Culex \rm mosquitoes, and egg development rate of \it Aedes \rm mosquitoes computed with the climate data for the region where cattle farm and markets  located in  the study area of Texas from January, $2005$ to October, $2010$, are shown in Fig. \ref{fig:b1andb3from05to10},  Fig. \ref{fig:theta3from05to10},  and Fig. \ref{fig:theta1from05to10}, respectively.   If the temperature is too low, the eggs will not develop into larvae and then adult mosquitoes. If the temperature is too high, the lifespan of the mosquitoes is shortened and the development rate decreases. Moisture index is the difference between precipitation and evaporation as shown in Equation (\ref{equation:Moisture}). A lower moisture index correlates to fewer adult mosquitoes because low moisture index represents a combination of low precipitation and high evaporation. For some days, the missing precipitation data from January, $2005$ to December, $2010$ \cite{NCDC2010} are assumed to be zero. The evaporation data are calculated using  Equation (\ref{equation:evaporation}) \cite{linacre1977simple}. The parameters in Equations (\ref{equation:birthrate}) through (\ref{equation:developmentrate}) are listed in Table \ref{table:culexparameters}. Although humans move between nodes, they do not transmit virus between nodes and the number of humans in each node (i.e., farm) is assumed to be fewer than $15$.

\allowdisplaybreaks
\begin{align}
b_1(Temp, precipitation, T_d,t)&=b_3(Temp, precipitation, T_d,t)=b_0+\frac{Emax}{1+exp[-\frac{Moisture(t)-Emean}{Evar}]}\label{equation:birthrate},\\
Moisture(t)&=\sum^t_{D=t-6}precipitation(D)-Evaporation(D) \label{equation:Moisture},\\
Evaporation(t)&=\frac{700(Temp(t)+0.006h)/(100-latitude)}{80-Temp(t)}\nonumber\\&+\frac{15(Temp(t)-T_d(t))}{80-Temp(t)}\label{equation:evaporation},\\
\theta_1(Temp,t)&=A_1*\frac{(Temp(t)+K)}{298.15}*\frac{exp[\frac{HA_1}{1.987}*(\frac{1}{298.15}-\frac{1}{Temp(t)+K})]}{1+exp[\frac{HH_1}{1.987}*(\frac{1}{TH_1}-\frac{1}{Temp(t)+K})]}\label{equation:theta1},\\
\theta_3(Temp,t)&=A*\frac{(Temp(t)+K)}{298.15}*\frac{exp[\frac{HA}{1.987}*(\frac{1}{298.15}-\frac{1}{Temp(t)+K})]}{1+exp[\frac{HH}{1.987}*(\frac{1}{TH}-\frac{1}{Temp(t)+K})]},\label{equation:developmentrate}
\end{align}
where\\
$Temp(t)=$air temperature in units of $^{o}C$ \cite{linacre1977simple}\\
$latitude=$the latitude (degrees) \cite{linacre1977simple}\\
$T_d(t)=$the mean dew-point in units of $^{o}C$ \cite{linacre1977simple}\\
$h=$the elevation (meters) \cite{linacre1977simple}\\
$K=$Kelvin parameter \\

\section*{Results}
\label{sec:model}
\subsection*{The Novel Mathematical Model}
Presented is a discrete time compartmental mathematical model based on a network approach. Rift Valley fever is transmitted by several species of mosquito vectors that have varying levels of vector competence; therefore, each genus and species combination requires modeling the vector competence, movement, and life stage development patterns which is too complicated while considering only a single species or genus is not accurate. Consequently, the species are loosely grouped as their genera and the parameters are allowed to vary following PERT distributions. The distribution captures uncertainties on inherent variability between species, as well as variability among individual mosquitoes. The mosquito parameters are functions of climate factors to reflect the impact of climate and season on mosquito dynamics. Only \it Aedes \rm and \it Culex \rm genera mosquitoes that are competent vectors of RVF virus transmission are considered in the model.

Different networks are developed for mosquito diffusion and livestock movement considering heterogeneity in both. In the cattle movement network, different types of nodes distinguish between sources, sinks, and transitions.

The model can be used to simulate networks with the number of nodes up to thousands with the easily solvable discrete time model. To use the model in any location, one only needs the initial populations, the movement rates, ranges of the parameters, and climate factors in each location to obtain the epidemic curve.

\subsection*{Case Study}
Sixteen initial conditions shown in Table \ref{table:16scenarios} in two regions of Texas, U.S.A from January $2005$ to October, $2010$ were tested with the model to determine their effects on the simulated and hypothetical spread of Rift Valley fever virus were it to be introduced. The average results of $100$ realizations for each scenario starting in the same small or large farm are presented qualitatively in Table \ref{table:scenarios}, and the quantitative numerical simulation results are shown in the Supporting Information section. For the simulations an introduction to a small farm is a farm with fewer than $10$ cattle and the large farm is considered a farm with more than $500$ cattle. By changing initial locations in extensive numerical simulations, we obtained different value for each variable from that of corresponding one in the table of Supporting Information but similar trends. Note at this time no specific mitigation strategies are applied here; during an outbreak the RVF virus control methods post detection will be expected to modify any such results.

\begin{table}[!ht]
\caption{
Sixteen different initial conditions. }
\centering
\begin{tabular}{|p{40pt}p{30pt}p{70pt}p{85pt}p{85pt}p{50pt}|}
\hline
\textbf {Farm size} &\textbf{Quantity} && &\bf{Infected }&\\
 && \it Aedes  \rm \bf{eggs}& \it Aedes  \rm \bf{mosquitoes}& \it Culex \rm \bf{mosquitoes}&\bf{Cattle}\\ \hline
{Small} &Few &\it Aedes  \rm-eggs-f-s &\it Aedes  \rm-f-s&\it Culex \rm-f-s &Cattle-f-s \\
&Many&\it Aedes  \rm-eggs-m-s &\it Aedes  \rm-m-s&\it Culex \rm-m-s &Cattle-m-s \\
{Large} &Few &\it Aedes  \rm-eggs-f-l &\it Aedes  \rm-f-l&\it Culex \rm-f-l &Cattle-f-l \\
&Many&\it Aedes  \rm-eggs-m-l &\it Aedes  \rm-m-l&\it Culex \rm-m-l &Cattle-m-l \\\hline
\end{tabular}
\label{table:16scenarios}
\end{table}

\subsubsection*{Size of the Epidemics}

The suffix –l or –s, (which denote large or small farms) were removed from the initial condition labels when comparing results with different initial infections in the same scale of initial location.  The impact of the Rift Valley fever epidemic in terms of infected cattle  depends on the size of the initial infection.

When the initial condition of the outbreak is assumed to be \it Aedes\rm-eggs-f (few \it Aedes \rm eggs), the simulations result in  a larger cumulative number of infected cattle than the one obtained in the case of \it Aedes\rm-eggs-m (many \it Aedes \rm eggs).  When the initial condition of the outbreak is assumed to be \it Aedes \rm-f (few adult \it Aedes \rm mosquitoes), the simulations result in a larger  cumulative number of infected cattle than the ones obtained in the case of \it Aedes\rm-m (many adult \it Aedes \rm mosquitoes). Similarly,  fewer initial infected \it Culex \rm mosquitoes  (\it Culex\rm-f)  leads to larger cumulative number of infected cattle than  the one obtained in the case of \it Culex\rm-m throughout the simulation period. 

When the initial condition of the outbreak is assumed to be Cattle-f (few cattle), the simulations result in a larger cumulative number of infected cattle than the ones obtained in the case of Cattle-m (many cattle).

The total number of infected humans and the total number of farms with at least one infected human remain fewer than one regardless of initial infection conditions. This is likely because the human population of each farm is assumed to be fewer than $15$. Therefore, human infection is unlikely in this case but this should not be inferred or generalized to be similar in a more heavily populated region or where there are many more persons in direct contact with animals (e.g., slaughter plants).
\subsubsection*{Timing of the Epidemics}
The temporal characteristics of Rift Valley fever cases followed the general trend that fewer infected individuals in the initial introduction resulted in a delayed epidemic peak. When the initial condition of the outbreak is assumed to be \it Aedes\rm-eggs-f-s, the simulation results in a peak $895$ days later than the one with initial starting conditions of \it Aedes\rm-eggs-m-s. When the initial condition of the outbreak is assumed to be \it Aedes\rm-eggs-f-l, the simulations result in a later peak than the \it Aedes\rm-eggs-m-l condition.  Comparing another pair of initial conditions, the epidemic peak happens no sooner when few initially infected \it Aedes \rm eggs are considered than when few initial infected \it Aedes \rm adult mosquitoes are assumed. Similarly, the epidemic peak happens not sooner when many initial infected \it Aedes \rm eggs are considered than the one when many initial infected \it Aedes \rm adult mosquitoes are assumed. When the initial condition of the outbreak is assumed to be \it Aedes\rm-f, the simulations result in a later peak than the \it Aedes\rm-l condition. When the initial condition of the outbreak is assumed to be \it Culex\rm-f, the simulations result in a later peak than the \it Culex\rm-l condition.  
 Few initially infected cattle produce a later peak than the one when many cattle are initially infected.

\section*{Discussion}
\label{sec:Conclusions}
The original meta-population model for Rift Valley fever described by Equations (\ref{equation:Aedeseggs}) through (\ref{equation:recoveredhuman}) has been proposed and applied to a case study in two study areas of Texas, U.S.A.  The simulation results are helpful in understanding the mechanisms of RVF virus transmission. Modeling each mosquito species individually requires  specific species information to parameterize the model, such as vector competence, which is often not available or is based on assumptions from other species. Therefore, the model groups competent mosquito vectors into two main genera of RVF competent mosquitoes, \it Aedes \rm and \it Culex\rm. The PERT distribution allows for mosquito species of the same genera to be clumped together and for individual variation within a single mosquito species by having a distribution with a most likely value and a range of possible values for each parameter. The distribution also allows the model to be easily adapted to new environments where the vector competence of mosquitoes remains uncharacterized. The model can accommodate various mosquito species of the same genus by adjusting the most likely values and the range of values to account for the variation in vector competence between species. Moreover, the model is not limited to the known mosquito vector species, and newly discovered competent vectors of RVF can be readily included in the model.

The model can be used to study not only local transmission between hosts and vectors, but also trans-location transmission of RVF virus with the network approach. The roles of mosquitoes and livestock in RVF virus transmission can be studied independently because they have separate networks. One infected farm node can spread the infection to other nodes connected to it; therefore, more nodes can be infected over time. The temporal and spatial evolution of RVF virus and its driving force can be analyzed. The spread of RVF virus is estimated within farms as well as between farms, markets, and feedlots. The goal of the simulation analysis is to provide insights into possible pathways for rapid spread of RVF virus among farms and counties. Using the cattle networks, the impact of cattle movement from trade can be investigated as newborn calves mature to weaning and on to harvest. The cattle farms are the source nodes where the cattle are born and raised for several months before being sold through markets or direct to feedlots, or to other farms as stockers or replacement females. Cattle  on an infected farm may become infected  and then carry the virus to the livestock market or else transition nodes before being sold to another farm, which may introduce the virus to a new farm. On the other hand, infected cattle movement to feedlots (sink nodes) does not propagate the transmission because there is no further transfer of cattle from the nodes except onto slaughter. Different mitigation strategies can be applied according to each node type (source, sink, and transition) within livestock movement network.

Discrete time modeling is  appealing in the way it describes the epidemic process, which is conceptualized as evolving through a set of discrete time epochs instead of continuously \cite{Longini1986}. Typically infections or illnesses are  reported at discrete time (daily or weekly)  \cite {Brauer2010, Longini1986}. Discrete time modeling makes it  easier to compare the incidence data with the output of  simulations \cite {Brauer2010}.  Moreover,   the numerical exploration of  discrete time models is more straightforward \cite {Brauer2010}. Thus, it can be easily implemented \cite{Brauer2010} by non-mathematicians \cite{KATRIEL2013, Brauer2010}, an advantage in the public health world \cite {Brauer2010}. Our  model allows for simulations of RVF outbreaks on small networks with a few nodes and large scale networks with thousands of nodes. The model is developed not only for the purpose of being applied to the study area of Texas, but also to any geographic region or habitat type of concerns without changing the model. To apply the model to a new study area, the modelers only need to adapt corresponding data into the model. It is time consuming and easy to make mistakes by frequently changing the model to adapt it to a new environment.

In large populations, with a large scale of epidemic incidence, deterministic models can provide good approximations \cite{Keeling2008}. Moreover, deterministic models are easier to analyze and interpret. However, the given starting condition and fixed parameters of a deterministic model will always result in the same solutions \cite{Ma2009} because deterministic models do not reflect the role of chances in disease spread \cite{Ma2009}. In principle, stochastic models are more realistic than deterministic models  in representing real world activities \cite{Keeling2008}. In a stochastic model, there are probabilities at each time step transferring from one epidemiological state to another \cite{Ma2009}. Hence, the outcomes of different runs may be different \cite{Ma2009} and a probability or credibility interval, similar to the confidence interval achieved from statistical analysis of empirical data, can be established. Stochastic models produce quantities such as the probability for an epidemic outbreak to occur and the mean epidemic duration time instead of  deterministic results \cite{Ma2009}. To reflect the chance of infection more appropriately, a stochastic model will later be developed. However, epidemic outcomes can still be compared with the presented deterministic model applied to case study in the study area of Texas, U.S.A.

Concerning the discussion of simulation results, \it Aedes \rm are the bridge between \it Culex \rm and livestock starting with  \it Aedes \rm egg infection. Infected \it Aedes \rm eggs may hatch infected \it Aedes \rm mosquitoes. The susceptible livestock become infected after being fed on by the infected \it Aedes \rm mosquitoes. \it Culex \rm mosquitoes are amplifiers of RVF virus transmission. \it Culex \rm mosquitoes acquire the infection after blood meals on infected livestock. In return, the infected \it Culex \rm feed on livestock and RVF virus infection is thus amplified. If there are more infected adult mosquitoes at the beginning, whether \it Aedes \rm or \it Culex \rm mosquitoes, the rate of infection is faster, herd immunity is reached faster, the cumulative number of infected cattle  is smaller because most recover before they further diffuse to other farms to spread RVF virus, as shown in Fig. \ref{fig:1and100CMLTsix11}. If most livestock infected by mosquitoes in a node recover before they move to other nodes, the number of infected livestock and mosquitoes that transmit RVF virus to other nodes  is reduced . The eggs do not hatch until their habitats, such as dambos (in Africa) or playas / ponds/ sloughs (Texas) are created by rainfall. Moreover, it takes time for \it Aedes \rm eggs to become adult \it Aedes \rm mosquitoes. Consequently, it may take longer to reach the epidemic peak with initially infected \it Aedes \rm eggs than with initially infected \it Aedes \rm mosquitoes.

\begin{figure}[h]
\centering
\includegraphics[angle=0,width=14cm,height=9cm]{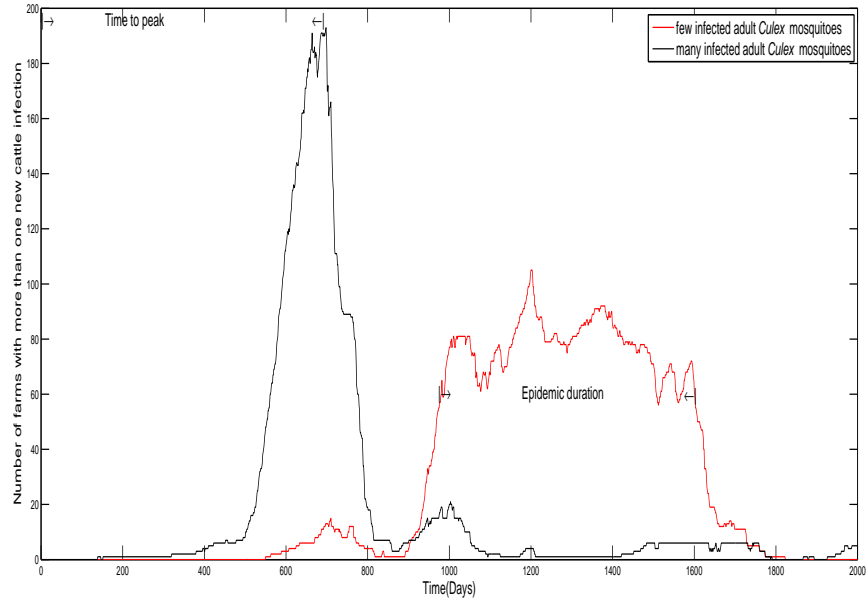}
\hspace{0.1in}
\caption{Disease epidemic characteristics based on model output with different initial numbers of infected \it Culex \rm     mosquitoes in a small farm.
 Time to peak infection is the time until the maximal number of cases is observed and epidemic duration is the amount of time an epidemic persists. }
\label{fig:Texas1and100CMS}
\end{figure}

Cattle can be spreaders of virus because they are frequently bought and sold \cite{Arino2007}. Infected cattle may infect a large number of mosquitoes via mosquito bites in a new location. In turn, the infected mosquitoes can bite a large number of susceptible cattle and transmit the virus to them. Movement bans during a RVF outbreak can restrict the further spatial spread of RVF. Therefore, very few infected cattle can infect a large number of susceptible cattle, by interacting with mosquito vectors. Early detection of infected cattle is essential. After local and regional authorities are warned and response planning initiated, such as cattle movement restrictions, culling, insecticide treatments, quarantines, and other methods to limit transmission can also be effective. These methods will be explored in future models.   The cumulative  number of infected cattle with few infected cattle at the beginning is larger than that with a large number of infected cattle at the beginning. The consequence caused by few initially infected  cattle should also be taken seriously.

There are no human cases (integers) in the simulations regardless of initial starting conditions because of the small  constant human population  in each node of the study region.   In  high population  areas,  there can be a large number of human cases.  Humans are often exposed to fewer mosquitoes than cattle,  especially in more developed countries, which results in lower probability of being infected by  mosquitoes. The probability that humans are infected by cattle  is also low in this region because the model does not account for contact with the virus via animal slaughter. Hence, the number of infected humans in each farm produced by simulations is fewer than $1$. Therefore, an introduction of RVF in the study area of Texas, U.S.A.  is likely to be mainly a concern for livestock farms and not an outbreak in humans as recently seen in South Africa based on the deterministic mathematical model presented by \cite{LingXue2010}. During previous outbreaks, many reported human cases proceeded with livestock cases. In the  U.S.A., humans still have the potential of being infected by mosquitoes and livestock especially when many livestock cases are reported. For this reason, the dynamics of human infection during an outbreak and the factors that affect RVF virus transmission will also be studied in future models.

In conclusion, the general epidemiological trend of a smaller initial infection observed through various simulations with various initial staring locations is: $(1)$ a larger total number of infected  cattle, $(2)$ a longer delay after introduction until the peak of the epidemic, and $(3)$ a more prolonged epidemic. If the infection remains small (and possibly undetected) for a longer duration, it expands geographically before the epidemic explodes involving many cattle almost simultaneously. Therefore, an established and endemic condition can generate larger epidemic disease incidence after a long period of apparent hibernation.




\begin{table}[!http]
\caption{
Parameters in Equations (\ref{equation:birthrate}) through (\ref{equation:developmentrate}). }
\centering
\begin{tabular}{|p{60pt} p{160pt} p{40pt} p{30pt}|}
\hline
\bf{Parameter} & \bf{Description} & \bf{Value} & \bf{Source}\\
\hline
$A_1$&parameter in Equation (\ref{equation:theta1}) & $0.15460$&\cite{Rueda1990}\\
$HA_1$&parameter in Equation (\ref{equation:theta1})& $33,255.57$ &\cite{Rueda1990}\\
$HH_1$&parameter in Equation (\ref{equation:theta1})&$50,543.49 $ &\cite{Rueda1990}\\
$TH_1$&parameter in Equation (\ref{equation:theta1}) & $301.67 $&\cite{Rueda1990}\\
$A$&parameter in Equation (\ref{equation:developmentrate}) & $0.25$&\cite{gong2010climate}\\
$HA$&parameter in Equation (\ref{equation:developmentrate})& $28094$ &\cite{gong2010climate}\\
$HH$&parameter in Equation (\ref{equation:developmentrate})&$35692 $ &\cite{gong2010climate}\\
$TH$&parameter in Equation (\ref{equation:developmentrate}) & $298.6$ &\cite{gong2010climate}\\
$b_0$&minimum constant fecundity rate &$0$ & \cite{gong2010climate}\\
$Emax$&maximum daily egg laying rate & $20$ &\cite{gong2010climate}\\
$Emean$&the mean of the daily egg laying rate & $0$& \cite{gong2010climate}\\
$Evar$& variance of function& $12$ & \cite{gong2010climate}\\
\hline
\end{tabular}
\label{table:culexparameters}
\end{table}

\begin{table}[!http]
\captionsetup{justification=justified, singlelinecheck=false}
\caption{
Qualitative numerical simulation results of different scenarios with respect to infected cattle. Numerical results are  in Table \ref{table:Quantitative}  in the appendix.  We define that if there is at least one cattle infected, then the farm is infected.  $A$ represents the number of infected farms. $B$ represents the cumulative number of infected cattle throughout simulation. $C$ is the total number of infected cattle when the number of infected cattle farms is maximum. $D$ denotes the time to peak  number of infected farms, that is, the time it takes from the first day to the day on which the largest number of infected farms appears as shown in Fig. \ref{fig:Texas1and100CMS}. $E$ denotes epidemic duration, defined as the number of days with more than $60$ infected cattle farms.
 The average number of infected farms in each day is in the range of $[350, 400)$, the average cumulative number of infected cattle during simulation is within the range $[350\times 10^3, 380\times 10^3)$, and the average time to peak is within $[1000, 1200)$.}

\centering
\begin{tabular}{|p{30pt} p{55pt} p{40pt}p{60pt}p{60pt}p{60pt} p{45pt}|}
\hline
 &&& \textbf{ Initial }&  \textbf{source } & \textbf{ of }& \textbf{infection}\\
\bf{Farm size} & \bf{Initial infection size} &\textbf{Outcome characteristics}&\it Aedes  \rm \bf{eggs}& \it Aedes  \rm \bf{ adult}& \it Culex \rm \bf{ adult}&\bf{Cattle}\\\hline
Small& Few ($1$) &$A$ &average &small &very small&very small \\
& & $B$	&very large	&very large	&large	&average\\
& & $C$	&very large	&very large	&average	&very small\\
&&$D$&very long&very long&long &medium \\
&& $E$ &medium&long&very long&short\\
&Many ($\gg 1$) &$A$  &very small &large&very large&average \\
&& $B$	&average	&small	&very small	&small\\
& & $C$	&very small	&small	&average	&very small\\
&&$D$ &short&short&short& short\\
&& $E$&short &very short &very short&very short\\
Large & Few ($1$)  & $A$&very small&very small&very small&small\\
&& $B$&very	large	&large	&average	&very large\\
&& $C$&very small &small &very small  &average \\
&& $D$& long& long&short &very long \\
&&  $E$ &very long&medium&short&long\\
& Many ($\gg 1$)  & $A$ &very large &very large&very large&very small\\
&& $B$&very small	&small	& small	&large\\
&& $C$&average& large&average &small \\
&& $D$ &short&very short&very short&long\\
&& $E$&very short& short&short& medium \\
\hline
\end{tabular}
\label{table:scenarios}
\end{table}

\section*{Acknowledgments}
We are grateful to Kenneth J. Linthicum for his suggestions on mosquito models, and we are thankful to  Bo Norby, Doyle Fuchs, Bryanna Pockrandt,  and Phillip Schumm for their help  in producing this work. We gratefully thank two  anonymous referees for their valuable comments and suggestions which lead to an improvement of our manuscript.
\bibliography{secondTexas}
\section*{Appendix}
The quantitative simulation results of different scenarios is shown in Table \ref{table:Quantitative}.

\begin{table}[!ht]
\captionsetup{justification=justified, singlelinecheck=false}
\caption{
Quantitative simulation results of different scenarios. We define that if there is at least one cattle infected, then the farm is infected. The number of infected  farms is represented by $A$ and the cumulative number of infected cattle throughout simulation is represented by $B$. The total number of infected cattle when the number of infected cattle farms is maximum ($C$). The time to peak number of infected farms ($D$)  means the time it takes from the first day to the day on which the largest number of infected farms appears.  Epidemic duration ($E$) means the number of days with more than $60$ infected cattle farms. The peak number of farms with more than one infected human  is represented by $F$ and the peak number of infected humans in a single farm in one day is represented by $G$. The total number of farms is $3526$ and 
the total number of cattle in all farms is $303240$.\\} 
\centering
\begin{tabular}{|p{30pt} p{50pt} p{50pt}p{60pt}p{70pt}p{60pt} p{40pt}|}
\hline
 &&& \textbf{ Initial}& \textbf{source} &\textbf{ of}& \bf{  infection}\\
\bf{Farm size} & \bf{Size of initial infection} &\textbf{Outcome characteristics}&\it Aedes  \rm \bf{eggs}& \bf{ Adult \it Aedes  \rm} &\bf{ Adult \it Culex  \rm  } &\bf{Cattle}\\\hline
Small&few&$A$ &$359$&$319$&$267$&$183$ \\
& & $B$ 	&$410 \times  10^3$&$411 \times  10^3$&$397 \times  10^3$&$374 \times  10^3$ \\
& & $C$ 	& $16288$&$6369$ &$4230$& $2557$\\
&&$D$&$1596$&$1382$&$1205$&$1012$  \\
&& $E$ &$444$&$471$&$592$&$291$ \\
&&$F$ &$0$&$0$&$0$& $0$\\
&& $G$&$0$ &$0$ &$0$ &$0$\\
&many &$A$  &$224$&$437$&$610$&$388$  \\
&& $B$&$364 \times  10^3$&$335 \times  10^3$&$313 \times  10^3$&$343 \times  10^3$  \\
& & $C$ 	&$1772$ &$3125$&$4433$&$2773$ \\
&&$D$ &$701$&$701$&$700$&$701$ \\
&& $E$&$278$&$181$&$217$&$227$ \\
&&$F$ &$0$&$0$&$0$& $0$\\
&& $G$&$0$&$0$&$0$&$0$\\
Large &few & $A$&$293$&$197$&$296$&$342$  \\
&& $B$&$407 \times  10^3$&$382 \times  10^3$&$354 \times  10^3$&$413 \times  10^3$ \\
& & $C$ 	& $2907$&$3878$&$2459$&$4411$ \\
&& $D$&$1205$&$1204$&$711$&$1382$ \\
&&  $E$&$557$&$443$&$278$&$467$ \\
&&$F$&$0$&$0$&$0$&$0$\\
&& $G$&$0$&$0$&$0$&$0$\\
&many & $A$ &$631$&$732$&$745$&$208$  \\
&& $B$&$315 \times  10^3$&$321 \times  10^3$&$332 \times  10^3$&$385 \times  10^3$ \\
& & $C$ 	&$4251$ &$4689$&$4428$&$3778$ \\
&& $D$ &$700$&$655$&$608$& $1204$\\
&& $E$&$226$&$260$&$276$&$449$ \\
&&$F$ &$0$&$0$&$0$&$0$\\
&& $G$&$0$&$0$&$0$&$0$\\
\hline
\end{tabular}
\label{table:Quantitative}
\end{table}

The outcome characteristics are classified into the following ranges.\\
very small ($0<A<300$  or  $0<B<320 \times  10^3 $ or $0<C<3000$),\\
small ($300\leqslant A<350$  or $320 \times  10^3 \leqslant B<350 \times  10^3$ or $  3000 \leqslant  C <4000$),\\
average ($350\leqslant A<400$ or  $350 \times  10^3 \leqslant B<380 \times  10^3$ or $  4000 \leqslant  C <4500$),\\
large ($400\leqslant A<600$ or $380 \times  10^3 \leqslant B<400 \times  10^3$ or $  4500 \leqslant  C <6000$),\\
very large ($A \geqslant 600$ or $B\geqslant 400 \times  10^3$ or $C \geqslant 6000$),\\
very short or really  large  $(0<D<700$ or $0<E<250)$,\\
short $(700\leqslant D<1000$ or $250 \leqslant E<300)$,\\
medium $(1000\leqslant D<1200$ or $300 \leqslant E<450)$,\\
long $(1200\leqslant D<1300$ or $450 \leqslant E<500 )$,\\
very long $(D\geqslant 1300$  or $ E\geqslant 500 )$.\\

\end{document}